\documentclass[article]{aa}
\usepackage[utf8]{inputenc}

\usepackage{graphicx}
\usepackage[]{natbib}
\usepackage{multirow}
\usepackage{amsmath} 
\usepackage{txfonts}  %% txfonts must be after amsmath!!!!!
\usepackage{enumitem}
\usepackage[colorlinks=true,linkcolor=blue,citecolor=blue]{hyperref}  %% disable for arxiv

\def\deg{$^{\circ}$}
\def\degb{^{\circ}}

\title{Possible evidence of ongoing planet formation in AB Aurigae\thanks{Based on data collected at the European Southern Observatory under programs 0104.C-0157, and 2104.C-5036}}

\subtitle{A showcase of the SPHERE/ALMA synergy}
\author{
A. Boccaletti\inst{\ref{lesia}}
\and E. Di Folco \inst{\ref{lab}}
\and E. Pantin \inst{\ref{cea}}
\and A. Dutrey\inst{\ref{lab}}
\and S. Guilloteau\inst{\ref{lab}}
\and Y. W. Tang\inst{\ref{taipei}}
\and V. Piétu \inst{\ref{iram}}
\and E. Habart\inst{\ref{ias}}
\and J. Milli\inst{\ref{ipag}}
\and T. L. Beck\inst{\ref{stsci}}
\and A.-L. Maire\inst{\ref{uliege}}
}
\institute{
LESIA, Observatoire de Paris, Universit{\'e} PSL, CNRS, Sorbonne Universit{\'e}, Univ. Paris Diderot, Sorbonne Paris Cit{\'e}, 5 place Jules Janssen, 92195 Meudon, France\label{lesia}
\and Laboratoire d'Astrophysique de Bordeaux, Université de Bordeaux, CNRS, B18N, Allée Geoffroy Saint-Hilaire, 33615, Pessac, France\label{lab}
\and 
 Laboratoire CEA, IRFU/DAp, AIM, Université Paris-Saclay, Université Paris Diderot, Sorbonne Paris Cité, CNRS, F-91191 Gif-sur-Yvette, France \label{cea}
\and Academia Sinica, Institute of Astronomy and Astrophysics, 11F of AS/NTU Astronomy-Mathematics Building, No.1, Sec. 4, Roosevelt Rd, Taipei, Taiwan \label{taipei}
\and IRAM, 300 rue de la piscine, Domaine Universitaire, 38406 Saint-Martin d'H\`eres, France\label{iram}
\and  Institut d'astrophysique spatiale, CNRS UMR 8617, Université Paris-Sud 11, Bât 121, 91405, Orsay, France\label{ias}
\and  Space Telescope Science Institute, 3700 San Martin Drive, Baltimore, MD, 21218, USA\label{stsci}
\and  CNRS, IPAG, Univ. Grenoble Alpes, F-38000 Grenoble, France\label{ipag}
\and STAR Institute, Université de Liège, Allée du Six Août 19c, B-4000
Liège, Belgium\label{uliege}
}
 \offprints{A. Boccaletti, \email{anthony.boccaletti@obspm.fr} }


\authorrunning{A. Boccaletti et al.}
\titlerunning{AB Aur}

\begin{document}

\abstract
        {Planet formation is expected to take place in the first million years of a planetary system through various processes, which remain to be tested through observations.  }
        {With the recent discovery, using ALMA,  of two gaseous spiral arms inside the $\sim$120\,au cavity and connected to dusty spirals, the famous protoplanetary disk around AB\,Aurigae presents a strong incentive for investigating the mechanisms that lead to giant planet formation. A candidate protoplanet located inside a spiral arm has already been claimed in an earlier study based on the same ALMA data.}
        {We used SPHERE at the Very Large Telescope (VLT) to perform near-infrared (IR) high-contrast imaging of AB\,Aur in
        polarized and unpolarized light in order to study the morphology of the disk and search for signs of planet formation.}
        {SPHERE has delivered the deepest images ever obtained for AB\,Aur in scattered light. Among the many structures that are yet to be understood, we identified not only the inner spiral arms, but we also resolved a feature in the form of a twist in the eastern spiral at a separation of about 30\,au. The twist of the spiral is perfectly reproduced with a planet-driven density wave model when projection effects are accounted for.
        We measured an azimuthal displacement with respect to the counterpart of this feature in the ALMA data, which is consistent with Keplerian motion on a 4-yr baseline.
        Another point sxce is detected near the edge of the inner ring, which is likely the result of scattering as opposed to the direct emission from a planet photosphere. We tentatively derived mass constraints for these two features.}
        {The twist and its apparent orbital motion could well be the first direct evidence of a connection between a protoplanet candidate and its manifestation as a spiral imprinted in the gas and dust distributions.} 
\maketitle
\begin{figure*}[ht!]
\begin{center}
\includegraphics[width=18.5cm]{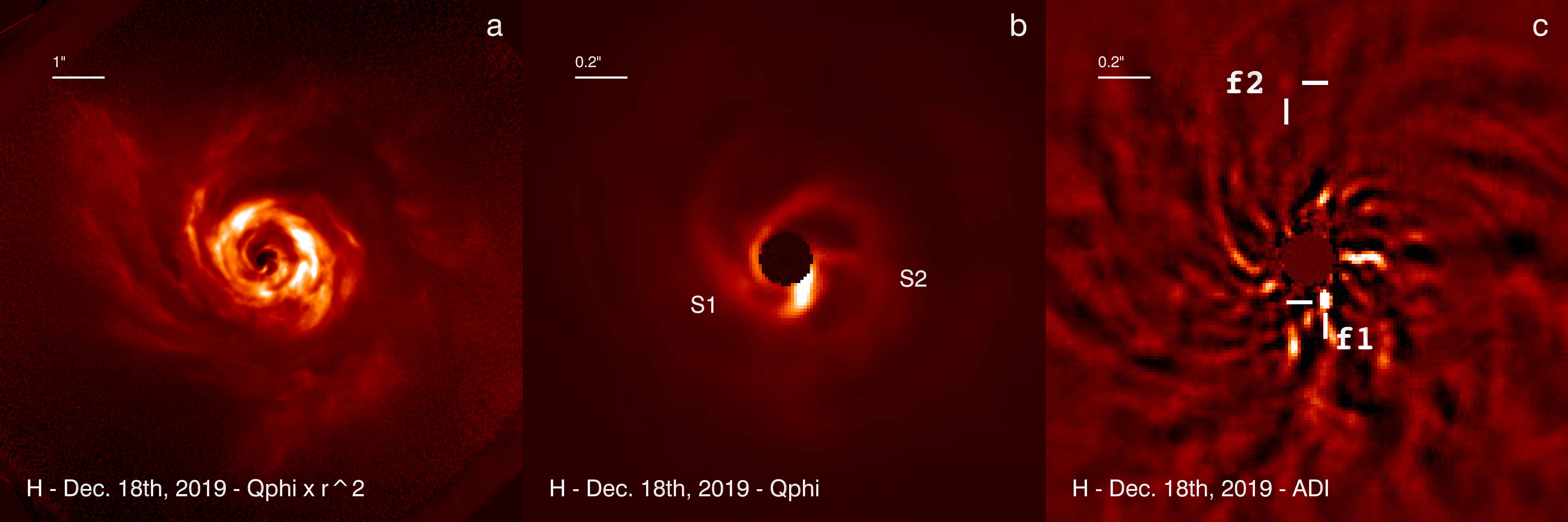}
\caption{Images of the AB Aur system obtained with SPHERE in polarized light (a, b) and unpolarized light (c). A large field of view ($10''$) is shown in panel a, where the polarized intensity has been multiplied with the square of the stellocentric distance ($Q_{\phi} \times r^2$) to visually enhance the outer part of the disk. A narrower field of view ($2''$) of the $Q_{\phi}$ map is displayed in panel b, in which the inner spirals are labelled S1 and S2. Comparable ADI-processed image is shown in panel c for total intensity in the H band filter. Features \texttt{f1} and \texttt{f2} are discussed in section \ref{sec:pointsources}. North is up, east is left }
\label{fig:general}
\end{center}
\end{figure*}

\section{Introduction}

%\LEt{ Regarding the title:\ A\&A avoids using questions in the titles of papers. I\ recommend revising to:\ Plausibility of ongoing planet formation in AB Aurigae or Possible evidence of ongoing planet formation in AB Aurigae}

Planets are believed to form in protoplanetary disks over a few million years. Images at all wavelengths exhibit a variety of structures, such as asymmetries, clumps, rings, or spirals. Most of these features, which hypothetically betray the presence of planets via gravitational or hydrodynamical effects, have been observed in disks orbiting Herbig Ae/Be stars because of their suitable brightness, particularly in the NIR \citep[see for instance][]{Grady2001, Clampin2003, Fukagawa2006, Isella2010, Christiaens2014}. Planet formation is generally favored \citep{Zhu2015} over instabilities in a massive self-gravitating disk \citep{Rice2005}.   In the early stage of planet formation, hydrodynamical simulations indicate that the accretion process generates at the planet location an inner and outer spiral pattern due to Lindblad resonances induced by disk-planet interactions \citep{Gressel2013}. While this crucial step is well-documented by theoretical works \citep{Dong2015, Bae2018}, observational evidence is rare and not fully conclusive. Recently, \cite{MuroArena2020} identified a kinked spiral in the protoplanetary disk SR\,21, suggesting that it may be a promising system for potentially witnessing ongoing planet formation.  

With a spectral type A0 \citep[$2.4\pm0.2M_\odot$,][]{DeWarf2003}, AB\,Aurigae is one of the closest \citep[$d=162.9\pm1.5$\,pc,][]{Gaia2018}, and one of the most intensively studied Herbig Ae star. Optical scattered light images of its protoplanetary disk have been obtained in the visible \citep{Grady1999, Fukagawa2004}, and in the near-IR \citep{Perrin2009, Hashimoto2011},  revealing a moderately inclined ($i\sim30\degb$), flared disk and prominent spiral patterns extending from 200 to 450\,au. In fact, AB Aurigae exhibits among the most spectacular spirals imaged so far in scattered light and high-contrast polarimetric imaging.

Millimetric observations have also been key in constraining the distribution of gas and dust in the AB\,Aur system. Using the IRAM Plateau de Bure Interferometer, \citet{Pietu2005} reported the presence of a large CO and dust rotating disk with a central cavity of an inner radius of $\sim 70$\,au. Surprisingly, \citet{Tang2012} found  counter-rotating CO spirals in the outer disk, which has been explained by projection effects of accretion flows arising from above the disk's midplane. The high accretion rate  \citep[$\sim 10^{-7}M_{\odot}$/yr,][]{Salyk2013} measured for this 1 Myr object reinforces this interpretation.

More recently, \citet{Tang2017} studied the CO 2-1 and dust continuum distribution at very high angular scale ($0.08''$ or 13\,au) using ALMA and revealed two spectacular CO spirals residing inside the large disk cavity. The authors speculated that these spirals are driven by planet formation and could be generated by two planets respectively located at $\sim$30 and 60-80\,au from the star. 
Motivated by these results, we  conducted VLT/SPHERE observations of the AB\,Aurigae system and present them in this letter. Here, we describe the observations, analyze the data and discuss the possible origin of the inner spiral pattern. 

%%%%%%%%%%%%%%%%%%%%%%%%%%%%%%%%%%%%%%%%%%%%%%%%%%%%%%%%%%%%%%%%%%%%%%
\section{Observations}
AB Aur (V=7.05, H=5.06, K=4.23) has been observed with SPHERE \citep{Beuzit2019} in polarimetry (Dec 2019) and in spectro-photometry  (Jan 2020).  The first data set was obtained with the infrared camera IRDIS \citep{Dohlen2008, deBoer2020} using Pupil Tracking (PT) in the H band, which allows us to retrieve, at the same time, the polarized intensity of the disk and the total intensity by making use of differential polarimetric imaging (DPI) and of angular differential imaging (ADI). The second epoch used the \texttt{IRDIFS-EXT} mode of SPHERE, which combines IRDIS in the K1K2 filters, and the spectrograph \citep[IFS,][]{Claudi2008} in low-resolution mode YJH (R=30), also processed with ADI. The observing log is provided in Table \ref{tab:log}. 

Both DPI and ADI increase the contrast by rejecting the starlight, exploiting either the fact that the stellar light is unpolarized or that off-axis objects rotate with the field in PT mode. While DPI provides unbiased polarimetric images of disks, ADI induces very strong photometric and astrometric artifacts owing to the self-subtraction effect \citep{Milli2012}. The data were processed at the SPHERE Data Center following \citet{Pavlov2008, Delorme2017, Maire2016}. The calibrated data cubes were then processed with SpeCal for total intensity data \citep{Galicher2018} and with a custom pipeline for polarimetric data \citep[as in][]{Bhowmik2019}. 
%The latter were compared to the output of the IRDAP pipeline \citep{vanHolstein2019} and found in good agreement. 
The total intensity data were reduced with Principal Component Analysis \citep{Soummer2012}, whereas for polarimetry, we calculated the azimuthal Stokes parameters $Q_\phi$ and $U_\phi$ maps following \citet{Schmid2006}. 
For a sanity check, we compared our polarimetric reductions with those of the  IRDAP pipeline \citep{vanHolstein2019}, which implements a careful treatment of the instrumental polarization. We found that since the instrumental effects are much fainter than the disk structures we further discuss below, the $Q_\phi$ images are identical in both pipelines.
The first epoch is significantly better in terms of starlight rejection, therefore the following sections are mainly based on these observations. 

\begin{figure*}[ht!]
\begin{center}
\includegraphics[width=18.5cm]{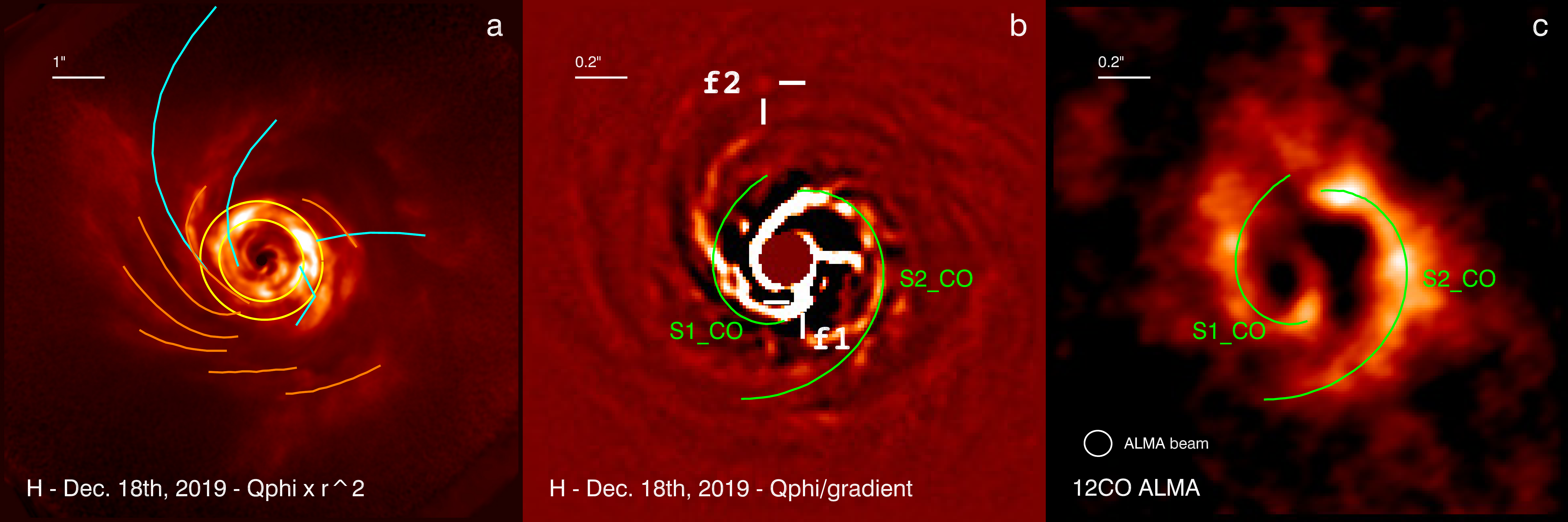}
\caption{Full-field polarimetric image (same as in Fig. \ref{fig:general}, $Q_{\phi} \times r^2$) displayed in panel a, on top of which several structures previously detected are overlaid: \citet{Hashimoto2011} spirals (orange lines), out of plane accreting spirals from \citet{Tang2012} (blue lines), and the dust ring (yellow ellipses) from the 0.9-mm ALMA image \citep{Tang2017}. The inner spiral arms (S1\_CO, S2\_CO) detected by \citet{Tang2017} are superimposed to the SPHERE polarimetric image $Q_{\phi}$, further processed with a gradient algorithm to enhance the inner spirals (b), and the ALMA CO map (c), in a $2''$ field of view. Features \texttt{f1} and \texttt{f2} are discussed in section \ref{sec:pointsources}. North is up, east is left.}
\label{fig:COspirals}
\end{center}
\end{figure*}

%%%%%%%%%%%%%%%%%%%%%%%%%%%%%%%%%%%%%%%%%%%%%%%%%%%%%%%%%%%%%%%%%%%%%%
\section{Disk morphology}
As the disk is moderately inclined, the ADI process strongly and irremediably attenuates  the disk signal (Fig. \ref{fig:general}c). Therefore, 
the following description is based on the polarimetric H-band $Q_{\phi}$ image (Fig. \ref{fig:general}a, b). 
The disk is visible all across the IRDIS $11'' \times 11''$ field of view. We  identify three main regions:  the dust ring, the outer spirals, and the inner spirals. 

The position of the ring detected with ALMA in the continuum \citep{Tang2017} is shown in Fig. \ref{fig:COspirals}a. In the sub-mm, it encompasses an elliptical region oriented at about 60$\degb$ north to east, from a minimal radius of $0.77''$ to a maximal separation of $1.19''$, corresponding to a ring of $\sim$140\,au on average (yellow ellipses in Fig. \ref{fig:COspirals}a). The aspect ratio is consistent with an inclination of 30$\degb$. In scattered light, the ring is not homogeneous with brighter parts in the west and the northeast, as opposed to the smooth 1.3\,mm image, which instead peaks to the southwest.  SPHERE resolves some structures in the ring that were already reported in scattered light by \citet{Oppenheimer2008} and \cite{Hashimoto2011}. In particular, a depleted region in the polarized flux, formerly reported by  \citet{Oppenheimer2008}, is also observed at PA=343$\degb$ although the contrast is low and its visibility is mostly a matter of intensity cut. We do not detect any point-source at this location, confirming the findings of \citet{Perrin2009}.
 
Outward of the ring, there are several spiral-like structures which, given the sensitivity achieved with SPHERE DPI, could in fact be the trace of one single grand spiral with several ramifications. This pattern likely starts  from the south and rolls clockwise while it gets broader with manifold branches (Fig. \ref{fig:general}a). It is particularly pronounced at the east and matches well to all the spirals identified in 2009 H-band data by \citet{Hashimoto2011} at Subaru (orange lines in Fig. \ref{fig:COspirals}a), indicating no perceptible motion in ten years, as already concluded by \citet{Lomax2016} on a shorter timescale. Therefore, this pattern is unlikely to be driven by planets inside the ring. Out-of-plane CO spirals \citep{Tang2012}, which have no obvious counterpart in scattered light, are indicated in cyan (Fig. \ref{fig:COspirals}a).

Inward of the ring ($<0.6''$), we detect two main spirals (S1 and S2 in Fig. \ref{fig:general}b), starting as close as the edge of the coronagraphic mask (about 95\,mas in radius) and in good agreement with the CO spirals identified by \citet[][and shown as green lines labeled S1\_CO and S2\_CO in Fig. \ref{fig:COspirals}b, c]{Tang2017}. The angular resolution brought by SPHERE evidences that the two inner spirals seem to intersect at the south of the star exactly at the location of an intensity peak ($\rho=0.21''$, $PA=177.7\degb$), which itself is found at a very similar position than the CO peak. S2 matches well the western CO spiral, while S1 seems to deviate from the eastern CO spiral. We suspect that this mismatch is due to the difference of angular resolution, which for ALMA is about $0.11''$, while SPHERE provides in H band a resolution of $0.04''$. As a consequence, it is difficult to disentangle S1 from S2 at the north in the CO map. S2 is more diffuse than S1 and appears larger in the west. High-pass filtering of the polarimetric image (Fig. \ref{fig:COspirals}b) shows higher complexity with clumps and several arms. 

While the CO map contains a cavity inward of the CO spirals the SPHERE observations reveal a third component ($PA\sim270\degb$ in Fig. \ref{fig:general}b) connecting the center of the image to S2, as a kind of "bridge" of $25-30$\,au long. This is not a post-processing artifact since this feature is seen both in the polarimetry image and in the total intensity ADI image, and was in fact already distinguishable in \citet{Hashimoto2011}, but  also in the gas line HCO$^+$ \citep{Riviere2019}. Whether it is a true connection or a projection effect cannot be determined with these observations alone. On each side of the "bridge," the polarized intensity is rather low, even lower than in between the spiral arms. Interestingly, in the direction of this feature the intensity of the spiral S2 and of the ring is fainter, possibly tracing a shadow pattern. If so, it will favor a feature near the midplane of the disk. Finally, S1 shows an increase of its width at $PA\sim115\degb$ as a compact clump that is $\sim0.12''$ long (Fig. \ref{fig:general}b, \ref{fig:COspirals}b), Complementary images are displayed in Appendix \ref{sec:complementdata}.

%%%%%%%%%%%%%%%%%%%%%%%%%%%%%%%%%%%%%%%%%%%%%%%%%%%%%%%%%%%%%%%%%%%%%%
\section{Point-like sources}
\label{sec:pointsources}
\begin{figure}[t]
\begin{center}
\includegraphics[width=9cm]{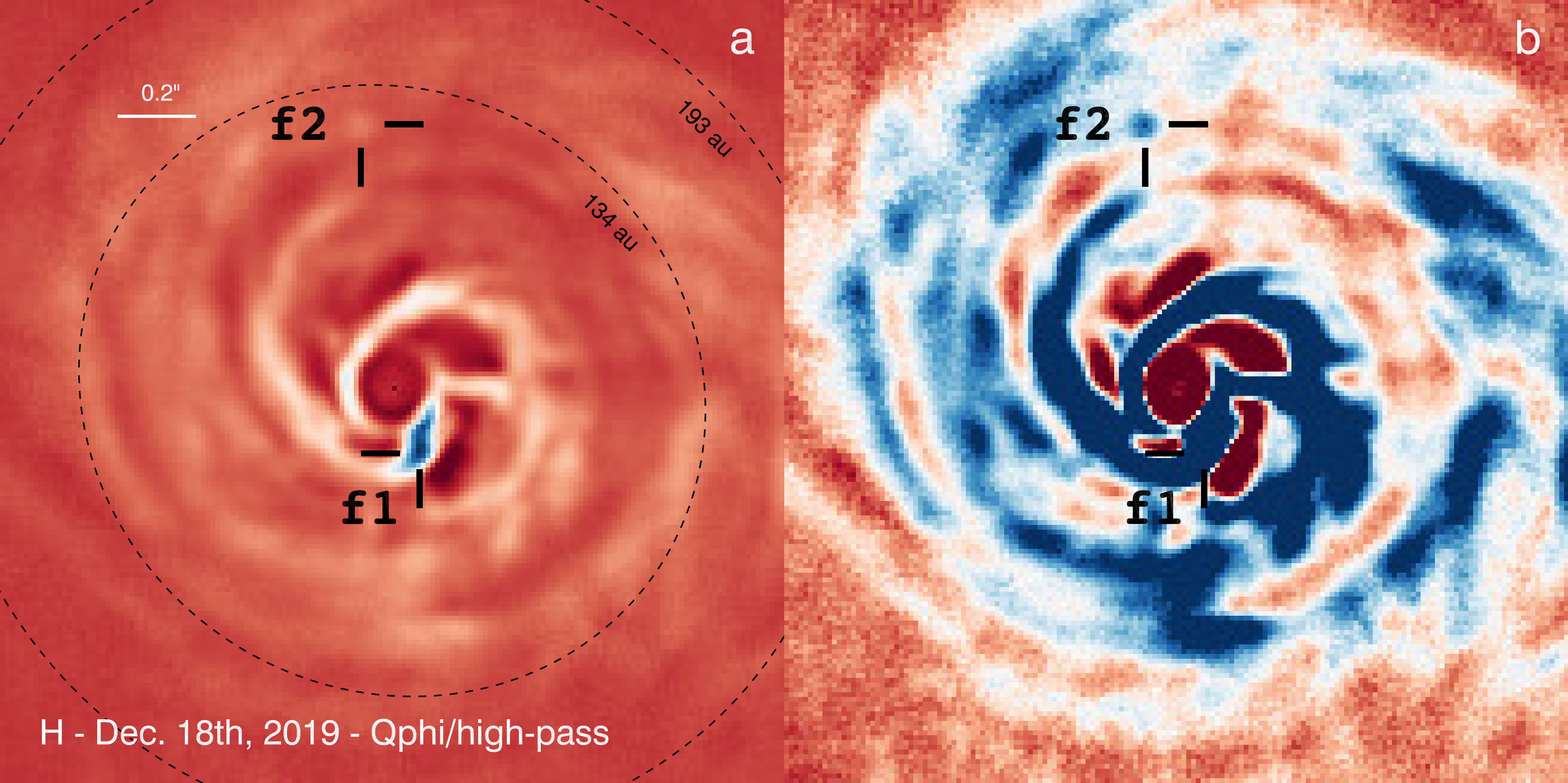}
\caption{Positions of features \texttt{f1} and \texttt{f2} overlaid on the polarimetric image shown for two intensity thresholds. The dusty ring is materialized with the dashed ellipses. For visualization purposes, the polarized intensity has been high-pass filtered and multiplied with the stellocentric distance ($Q_{\phi}\times r$).  }
\label{fig:cc}
\end{center}
\end{figure}

Many structures are observed in the disk but two features have a strong interest. The first one (\texttt{f1}) is the strongest signal in Fig. \ref{fig:general}b and c, south of the coronagraphic mask edge, and a second one (\texttt{f2}) is a nearly point source further out to the north. Interestingly, both show up in DPI and ADI images (Fig. \ref{fig:cc}). They are also detected at the second epoch in total intensity (ADI), but with a much lower signal to noise ratio (Fig. \ref{fig:appendix}). 

We extracted the photometry and astrometry with SpeCal.  The positions of the sources are based on the ADI reduction  using the PCA 5 modes algorithm (Fig. \ref{fig:COspirals}b) and the negative fake planet injection method \citep[for details see][]{Galicher2018}, but the values are consistent within error bars for all algorithms. \texttt{f1} is located at $\rho=0.160\pm0.006''$, $PA= 203.9\pm2.9\degb$. It coincides with the root of the eastern spiral identified as S1 in the ALMA image, where the two spirals S1 and S2 observed with SPHERE appear to intersect. The shape of this feature in the $Q_{\phi}$ image (Fig. \ref{fig:cc}a) is remarkable and we discuss it further in the next section. The second source, \texttt{f2}, lies at the very edge of the dust ring inside the cavity,  in the prolongation of the eastern spiral. Its position is $\rho=0.681\pm0.006''$, $PA= 7.6\pm1.8\degb$.  

\texttt{Feature f1} is elongated while \texttt{f2} is nearly point-like. This characteristic is confirmed by the photometric analysis. The contrast relative to the star derived from IRDIS data in the H band and measured at \texttt{f1} is quite variable from one ADI reduction to another, ranging from $6.5\pm0.8\times10^{-4}$ to $6.8\pm0.6\times10^{-5}$, depending on the effectiveness (ability to suppress the starlight) of the ADI algorithm. 

Similarly, the photometry in the K1K2 IRDIS image  obtained at the second epoch varies from one ADI reduction to the other, but with a small difference of contrast between the two narrow band filters: $3.4\pm0.5\times10^{-4}$ and $4.3\pm0.5\times10^{-4}$ , respectively. Whether it is indicative of a red color remains to be confirmed with higher quality data in the K band. Overall, we cannot consider the total intensity contrast measurements of \texttt{f1} reliable as it cannot be assimilated to a point source. On the other hand, \texttt{f2} has a stable contrast in the H band of $4.9\pm0.5\times10^{-6}$, and $4.8\pm0.8\times10^{-6}$ in K1K2, indicative of a bona-fide point source, but with a grey color.
Photometric measurements with the IFS (not shown here) in the YJH bands are not much more conclusive regarding \texttt{f1}, while \texttt{f2} is actually not detected. 

The fact that \texttt{f1} and \texttt{f2} are detectable in polarimetry and total intensity strongly suggests that these signals cannot result from the  photosphere of some protoplanets. Instead, they are likely to be the telltale of dust particles detected from scattering or from other thermally induced mechanisms. 

%%%%%%%%%%%%%%%%%%%%%%%%%%%%%%%%%%%%%%%%%%%%%%%%%%%%%%%%%%%%%%%%%%%%%%
\section{Spirals}
\label{sec:spirals}

The presence of spiral arms in protoplanetary disks has been usually associated to the indirect clues of planets provided that the disks are not massive enough to undergo self-gravity \citep[][for instance]{Muto2012, Boccaletti2013, Benisty2015}. This link with a gravitational perturber has never been definitely established, except for the prominent spiral arms in HD\,100453, very likely launched by a stellar M-dwarf companion \citep{Dong2016}. In that respect, the ALMA observations of AB Aur \citep{Tang2017} are likely the only case, together with GG Tau A \citep{Phuong2020}, to provide a strong suspicion for the presence of a protoplanet associated to spiral arms. 

To model the morphology of the inner spirals we rely on the formulation of \citet{Muto2012} described in Appendix \ref{sec:muto}. A direct fit of this analytical relation (Eq. \ref{eq:muto}) onto the spiral positions was found too  unstable given the difficulty to isolate the spine of the spiral arms against the other disk structures. Instead, we proceed with a manual adjustment of the theoretical spiral location taking into account the disk inclination ($i=30\degb$), the disk Position Angle ($PA=60\degb$), the 
gas disk vertical scale-height ($H_0=25$\,au at 100\,au) and the flaring index ($r^{1.2}$), following \citet{diFolco2009, Hashimoto2011, Tang2012, Li2016}. 
Because the spirals are pressure bumps and the near IR emission is optically thick, the net effect is a local variation of the disk thickness, hence these spirals materialize at the disk surface and not in the mid-plane \citep{Juhasz2015}. Therefore, we assume that the altitude of the disk layer at which the optical depth ($\tau_{\nu}$) is equal to one is $H(\tau=1)=1.5\times H_0$. These values are used to project the analytical relation $\theta(r)$ onto the plane of the sky and are known to some level of accuracy. Therefore, the solutions presented here provide qualitative results, which may not be unique. We note, in particular, a degeneracy between the inclination, the disk scale-height, and the radial distance of the putative planet, whereas the flaring index has only a weak impact on the shape of the projected spiral.

\begin{figure*}[t]
\begin{center}
\includegraphics[width=18.5cm]{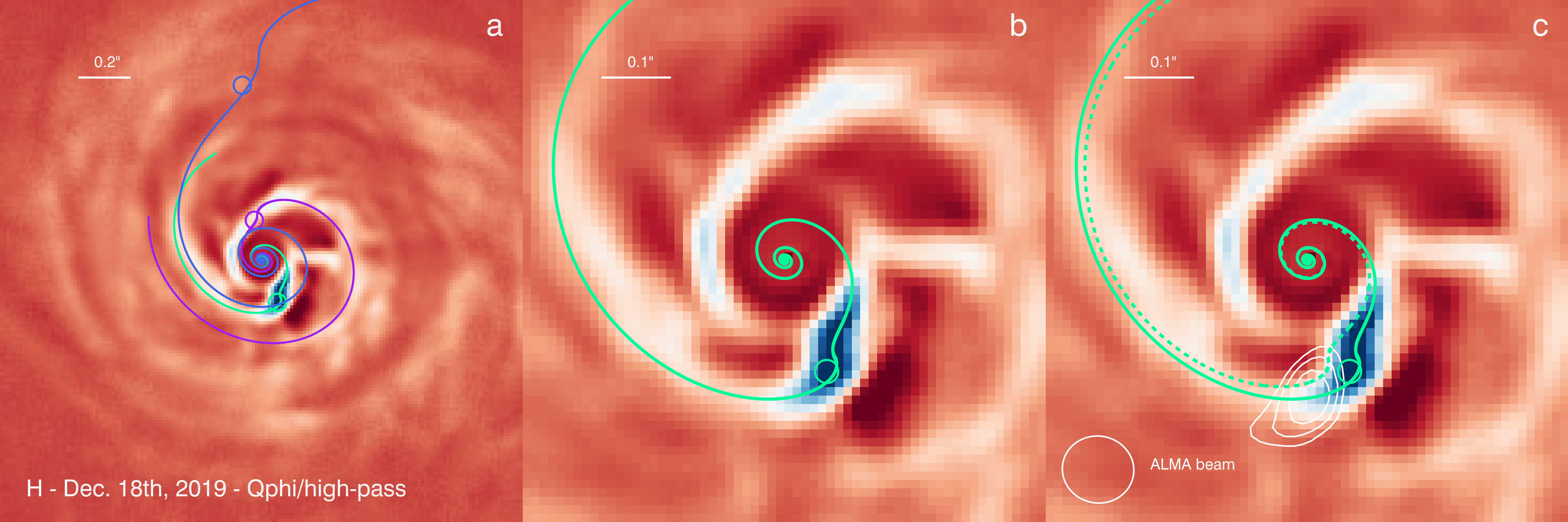}
\caption{Same data as Fig. \ref{fig:cc} with three different models of spirals matching S1 (green and blue lines) and S2 (purple line). The positions of the perturbers are indicated with a circle (panel a). Panel b is a zoom in version of panel a, where only the best model reproducing the spiral twist in S1 is shown. 
In panel c, the contours of the CO map at the location of the candidate protoplanet are overlaid, together with the spiral model offset by $14.1\degb$ (dashed green line), corresponding to the 4-yr timelapse between ALMA and SPHERE observations.}
\label{fig:spirals}
\end{center}
\end{figure*}

To account for S1 we considered two cases for which we forced a planet to be located close to the positions of the features discussed in section \ref{sec:pointsources}. A planet located at $r_c=0.184''$, $\theta_0=196.8\degb$ (in the orbital plane of the disk) reproduces the shape of the spiral and is also in accordance with the location of \texttt{f1} as seen in Fig. \ref{fig:spirals}a, and b (green line). The analytical model of Eq. \ref{eq:muto}, while simple, matches the "S" shape expected for a spiral density wave surprisingly well as it goes on to trigger the accretion of gas onto a protoplanet. On the contrary, imposing the planet location at \texttt{f2} matches only the external part of the spiral S1 (blue line in Fig. \ref{fig:spirals}a). For comparison, the contours of the CO peak in the ALMA image is shown in Fig. \ref{fig:spirals}b. 
Being localized at the surface, the spiral twists appear slightly offset from the perturber positions (circles), which are in the midplane. 

For the western spiral S2, since there is no obvious features nor strong signal which we can consider as a source, we blindly searched for the best matching model. The purple line in Fig. \ref{fig:spirals}a corresponds to  $r_c=0.172''$, $\theta_0=6.0\degb$. This model reproduces well the external part of the western spiral, which presents a more diffuse aspect or broader shape than the eastern one, but it fails to match the innermost, eastern arm which is less tightly wound as it approaches the occulting spot ($0.1'' < r < 0.16''$).

%%%%%%%%%%%%%%%%%%%%%%%%%%%%%%%%%%%%%%%%%%%%%%%%%%%%%%%%%%%%%%%%%%%%%%
\section{Discussions}
We provide a further exploratory analysis of the two identified objects, noting that they have both been detected in polarimetry and total intensity. 

The object labeled \texttt{f1} is fitting the source of the S1 spiral which at this location features an "S" shape,  suggestive of the stationary pattern composed of inward and outward spiral density waves excited by the gravitational potential of an accreting protoplanet \citep[see ][for a review]{Kley2012}. The spirals are trailing, as predicted by theory, since the near side is the southeast part of the disk \citep{Fukagawa2004} and the  resolved velocity field \citep{Pietu2005} implies that the disk  rotates counterclockwise. However, it is  not a straightforward exercise to determine whether the signal is from pure scattering (polarized) or a mix of scattering and emission (usually unpolarized).  \citet{Zhu2015} has calculated the spectral energy distribution emitted from an accreting circumplanetary disk, which can be compared to our photometric measurements, assuming the signal of \texttt{f1} is dominated by emission. The luminosity varies as the product of the planet mass with the accretion rate, and also depends on the inner radius of the circumplanetary disk. These three parameters necessarily lead to a high degree of degeneracy.
Although it is difficult to overcome the ADI bias in total intensity measured for the particular case of \texttt{f1}, as it is not point-like, we measured a minimum contrast of $6.5\times10^{-4}$ in the H band. If we assume a moderate accretion rate of $10^{-8}M_{\odot}$/yr as in \citet{Zhu2015}, it translates into a planet mass of $4 - 13 M_J$. We note that the accretion rates estimated for PDS 70b are orders of magnitude lower \citep{Haffert2019, Hashimoto2020}, which then would imply stellar mass for \texttt{f1}. As a consequence, the flux of \texttt{f1} is likely not to be dominated by thermal emission.   

On the contrary, the object \texttt{f2} is not directly associated to a spiral, so we assume its flux can be attributed to pure scattering. Mass constraints can be derived from dynamical argument. Indeed, the inner edge of the disk cavity at $\sim140$\,au can be sculpted by the chaotic zone of a planetary mass object. Assuming a circular orbit for the sake of simplicity, \citet{Wisdom1980} provides a relation between the planet/star mass ratio, and the inner edge/planet distance ratio, from which we derived a mass of $\sim 3 M_J$. The AB Aur cavity is very large, such that, if resulting from dynamical interplay with planets, several bodies including \texttt{f1} should be involved.

Finally, the object \texttt{f1} appears rotated counterclockwise from the CO peak in the ALMA image which, considering the stellar mass, the 4-yr interval, and the physical stellocentric distance of \texttt{f1}, is in excellent agreement with the expected rotation ($14.1\pm1.1\degb$ in the orbital plane) of the spiral due to the Keplerian motion of the protoplanet candidate (Fig. \ref{fig:spirals}c). However, the peak of the CO emission is
at a radial distance slightly larger than expected. There are a few technical reasons accounting for this difference, which remain to be investigated thoroughly. 
Nevertheless, to our knowledge, this is the first time that we can confidently associate the rotation of a spiral pattern with the orbital motion of a planet candidate, which reinforces our confidence in the nature of the detected spiral twist. 

In conclusion, the SPHERE observations of AB\,Aur in scattered light combined to the ALMA data in the thermal regime provide strong evidence that we are actually witnessing ongoing planet formation revealed by its associated spiral arms. Further observations would be required to confirm this result and to derive better mass estimates for  potential planets in this location.  

%%%%%%%%%%%%%%%%%%%%%%%%%%%%%%%%%%%%%%%%%%%%%%%%%%%%%%%%%%%%%%%%%%%%%%

%%%%%%%%%%%%%%%%%%%%%%%%%%%%%%%%%%%%%%%%%%%%%%%%%%%%%%%%%%%%%%%%%%%%%%
\begin{acknowledgement}
% french support
French co-authors acknowledge financial support from the Programme National de Planétologie (PNP) and the Programme National de Physique Stellaire (PNPS) of CNRS-INSU in France. 
% DC
This work has made use of the SPHERE Data Centre, jointly operated by OSUG/IPAG (Grenoble), PYTHEAS/LAM/CESAM (Marseille), OCA/Lagrange (Nice) and Observatoire de Paris/LESIA (Paris). We thank P. Delorme (SPHERE Data Centre) for his efficient help during the data reduction process. 
YWT acknowledges support through MoST grant 108-2112-M-001-004-MY2. 
ALM acknowledges the financial support of the F.R.S-FNRS through a postdoctoral researcher grant.
\end{acknowledgement}

\bibliography{abaur}

\begin{appendix}
\section{Complementary data}
\label{sec:complementdata}

%----------------------------
\begin{table*}[th!]
\begin{center}
\begin{tabular}{l l r l r l r l r l r l r l r l r }
\hline
Date UT               &       Filter                  &         Fov rotation    &       DIT     &       N$_ \mathrm{exp}$       &       T$_\mathrm{exp}$        &     seeing                      &       $\tau_0$      & Flux var.  &       TN                      \\
                        &                                                  &       (\deg)          &       (s)     &                                       &       (s)                             &       ($''$)                          &       (ms)    & (\%)   &    (\deg)                 \\ \hline \hline
2019-12-18  &  IRDIS - H & 29.6     &   32  &   160 &    5120 &  $0.67\pm0.09$   &   $7.1\pm0.9$ & $\sim$2 &    $-1.70$\\
\hline
2020-01-17  &  IRDIS - K1K2    &  38.5    &   16  &   256 &   4096  &  $0.71\pm0.22$   &   $5.6\pm1.1$ & $\sim$2  &    $-1.77 $ \\
2020-01-17  &  IFS - YJH     &  38.5    &   64  &   64  &   4096 &   $0.71\pm0.22$   &   $5.6\pm1.1$ & $\sim$2  &    $-1.77 $\\\hline

\hline
\end{tabular}
\end{center}
\caption{Log of SPHERE observations indicating (left to right columns): the date of observations in UT, the filters combination, the amount of field rotation in degrees, the individual exposure time (DIT) in seconds, the total number of exposures, the total exposure time in seconds, the DIMM seeing measured in arcseconds, the correlation time $\tau_0$ in milliseconds, the variation of the flux during the sequence in \%, and the true north (TN) offset in degrees. } 
\label{tab:log}
\end{table*}
%----------------------------

\begin{figure*}[t]
\begin{center}
\includegraphics[width=18.5cm]{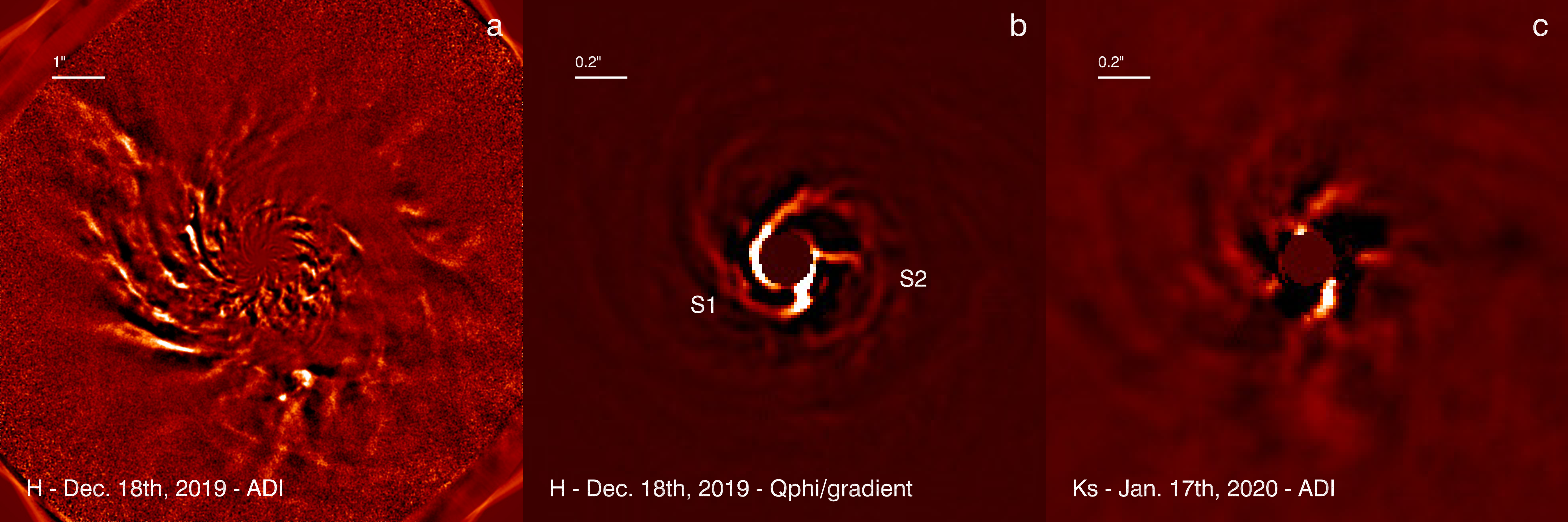}
\caption{
Complementary images of the AB Aur system.
(a) ADI image (H band) in a large field of view ($10''$) multiplied with the square of the stellocentric distance. (b) high-pass filtering (gradient) of the DPI image to enhance the spirals S1 and S2. (c) second epoch ADI observations in the K1K2 filters.}
\label{fig:appendix}
\end{center}
\end{figure*}

\section{Expression of a spiral arm}
\label{sec:muto}

According to the theory of density waves generated by a gravitational perturber \citep{Rafikov2002} in protoplanetary disks, \citet{Muto2012} proposed an approximate linear relation to describe the shape of a spiral $\theta(r)$, as a function of the planet location ($r_c$, $\theta_0$) and disk properties, as follows:

\begin{eqnarray}
 \theta(r) &=& \theta_0-\frac{\text{sgn}(r-r_c)}{h_c}       \nonumber \\
        &&\times  \left[  \left(\frac{r}{r_c}\right)^{1+\beta} \left\{ \frac{1}{1+\beta}-\frac{1}{1-\alpha+\beta} \left(\frac{r}{r_c}\right)^{-\alpha}  \right\} \right. \nonumber\\
   && \left. -\left( \frac{1}{1+\beta} - \frac{1}{1-\alpha+\beta} \right)\right]
\label{eq:muto}
,\end{eqnarray}

where $h_c$ is the disk aspect ratio at the planet location, $\alpha$ and $\beta$ are the power-law exponents of the angular frequency and temperature profile dependence with $r$. 
In this paper, we assumed $\alpha=1.5$ for the Keplerian velocity and $\beta=0.25$ to account for the temperature profile which varies as $r^{-2\beta}$.

This formulation is now commonly used to interpret high-contrast images in which spiral patterns are observed, but it involves a number of assumptions, in particular, the idea that small grains observed in scattered light are coupled to the gas and that a single planet induces a single spiral arm. Clearly, this is  not always  supported by hydro-dynamical simulations, which result in much more complex morphologies. In a more realistic situation, a planet could produce several spiral arms \citep{Dong2015} or planet formation could inevitably lead  to multiple cores and, hence, to manifold spirals \citep{Kadam2019}.

\end{appendix}

\end{document}